\documentclass[12pt]{article}
\usepackage{a4wide,color}
\usepackage{cite,amstext,amsfonts,amsmath}
\font\mybb=msbm10 at 10pt
\def\bb#1{\hbox{\mybb#1}}
\def\be{\begin{equation}}
\def\ee{\end{equation}}
\def\bseq{\begin{subequations}}
\def\eseq{\end{subequations}}

\def\bea{\begin{eqnarray}}
\def\eea{\end{eqnarray}}

\def\bseq{\begin{subequations}}
\def\eseq{\end{subequations}}

\begin{document}

\begin{flushright}
V1. June 23, 2019. V2. November 19, 2019
\end{flushright}

\vspace{2cm}

\begin{center}
{\LARGE
Superstring at the boundary of open supermembrane interacting with D=4 supergravity and matter supermultiplets }

\vspace{1cm}

{\large\bf Igor Bandos}
\\
\vspace{1cm}
{\small\it Department of
Theoretical Physics, University of the Basque Country UPV/EHU, \\ P.O. Box 644, 48080 Bilbao, Spain} \\
{\small\it and IKERBASQUE, Basque Foundation for Science, 48011, Bilbao, Spain}

\vspace{2cm}

\thispagestyle{empty}

{\bf Abstract}
\end{center}

We present the complete supersymmetric and $\kappa$--symmetric action for the 4-dimensional interacting system of   open  supermembrane, dynamical supergravity and  3--form matter multiplets. The cases of a single 3-form matter multiplet and a quite generic model with a number of nonlinear interacting double 3-form multiplets are considered.
In all cases the fermionic parameter of the $\kappa$-symmetry
is subject to two apparently different  projection conditions which suggests that the ground state of the system, in particular a domain junction,
might preserve at most 1/4 of the spacetime supersymmetry.

The boundary term of the open supermembrane action, needed to preserve the $\kappa$-symmetry, has the meaning of the action of a superstring.
The Wess--Zumino term of this superstring action is expressed in terms of real linear superfield playing the role of St\"uckelberg field for the 3--form gauge symmetry. This indicates that this symmetry is broken spontaneously by the superstring at the boundary  of supermembrane.

\vspace{3.0cm}

\tableofcontents

\setcounter{equation}0

\section{Introduction}

Eleven dimensional supermembrane \cite{Bergshoeff:1987cm,Bergshoeff:1987qx}, presently also known under the name of M2-brane, is one of the most important fundamental objects of the hypothetical underlying M-theory. Its consistency in curved superspace background subject this to the equations of motion of the 11-dimensional supergravity \cite{Cremmer:1978km,Cremmer:1980ru,Brink:1980az}, which is believed to provide the low energy limit of the M-theory.

The simpler 4D cousin of M2-brane also attracted an interest already in late 80-th \cite{Achucarro:1988qb}. Different aspects of its interaction with ${\cal N}=1$ $D=4$ supergravity and matter multiplets were the subject of study in
\cite{Ovrut:1997ur,Huebscher:2009bp,Bandos:2010yy,Bandos:2011fw,Bandos:2012da,Bandos:2012gz,Kuzenko:2017vil,Bandos:2018gjp,Bandos:2019qok}. The selfconsistency of nontrivial interaction with supermembrane requires matter and supergravity supermultiplets to include three form fields, thus leading naturally to the so-called variant superfield representations
\cite{Gates:1980ay,Gates:1980az,Binetruy:1996xw,Ovrut:1997ur,Farrar:1997fn,Kuzenko:2005wh,Bandos:2010yy,Bandos:2011fw,Bandos:2012da,Bandos:2012gz,Kuzenko:2017vil,Farakos:2016hly,Farakos:2017jme}.

In particular the interaction of closed supermembrane with supergravity and matter in the models of the type appearing in string theory  compactifications was studied in  \cite{Bandos:2018gjp} while  \cite{Bandos:2019qok} considered supermembrane interaction with supersymmetric $SU(N)$ Yang-Mills ($SU(N)$ SYM)  theory and its effective description by Veneziano-Yankelovich (VY)  action \cite{Veneziano:1982ah}. In the latter case the accounting for the presence of  supermembrane allowed to solve a long-standing problem \cite{Kogan:1997dt} of the missing contribution to the tension of BPS saturated  domain--wall  configurations,  for  which  the  membrane  serves  as  a core. This also allowed us to find in \cite{Bandos:2019qok} the explicit BPS domain wall solutions in this theory.

In \cite{Bandos:2019qok} also the interacting action for  open supermembrane carrying string on its boundary and rigid supersymmetric theories: generalized Wess-Zumino models including VY/SYM model,  was briefly discussed.
To our best knowledge the generic case of D=4 interacting system of open supermembrane, superstring on its end, supergravity and p-form matter has not been studied yet. The aim of this paper is to create a basis to feel this gap\footnote{The actions for open  M2 brane (D=11 supermembrane) ending on M9-brane (the Horava-Witten end-of-world nine-brane) and M5-brane were found in  \cite{Cederwall:1997hg,Brax:1997ka} and \cite{Brax:1997ht}. However, as the off-shell superfield description of the 11D supergravity and 10D matter was not known (and is still unknown), the   11D supergravity and 10D matter ($E_8$ SYM) at the boundary of 11D spacetime were considered as a background obeying the 'free' equations of motion without superbrane sources.
Probably a bypass  to the complete but gauge fixed Lagrangian description of the dynamical system including, besides open 11D supermembrane, also 11D supergravity and 10D matter, can be reached on the line of \cite{Bandos:2001jx,Bandos:2002kk,Bandos:2005ww}. }.  We present the  complete superfield action for such an interacting system and prove its $\kappa$-symmetry which is an important property indicating that  the ground state of this dynamical system preserves a part of supersymmetry (and hance is a stable BPS state). The complete interacting action can be split on the sum of the  terms describing  supergravity plus matter system ($S_{sugra+matter}$), supermembrane ($S_{p=2}$) and  superstring at the end of open supermembrane ($S_{p=1}$)
\be\label{S=S+}
S= S_{sugra+matter}+ S_{p=2}+S_{p=1}\; .
\ee
We will describe these ingredients step by step for the case of coupling to different formulations of supergravity and different types of matter system. Clearly, $S_{p=2}$ and $S_{p=1}$ can be treated as actions of supermembrane and superstring at the end of supermembrane in the background of supergravity and matter multiplets.

Our action can be used to describe an effective field theories of string compactifications with open branes and branes at the boundary of open branes \footnote{See for instance \cite{Kachru:2003sx,Baumann:2007ah,Moritz:2017xto} and  \cite{Cribiori:2019hod} and refs. therein for the description of a (not always direct) way from  10D branes and supergravity to 4D effective theories. } and to study  the role of open and intersecting supermembranes in supersymmetric generalizations and/or deformations  of the constructions  from  \cite{Dvali:2005an,Groh:2012tf,Almeida:2019xzt,Font:2019cxq,Lee:2019xtm}. It will be interesting to search for the supersymmetric domain wall junction solutions (see \cite{Ritz:2004mp,Shifman:2009zz}) of the equations of motion with open membrane sources which follow from our interacting actions.


In string theory the system of 4D open supermembranes with superstrings at the boundary of their wordlvolume
 can be obtained  from the network of higher p-branes on flux vacua of the type considered in \cite{Evslin:2007ti} and \cite{BerasaluceGonzalez:2012zn}.
In particular, the systems  of connected domain walls and strings which appear from networks of D7-,  D5- and D3-branes  in compactifications on wrapped Calabi-Yau manifolds are described in the approach of calibrations in \cite{Evslin:2007ti} were the explicit examples in the case of  toroidal orientifold vacua and the Klebanov-Strassler geometry  \cite{Klebanov:2000hb} have been discussed in more detail.
In \cite{BerasaluceGonzalez:2012zn} the authors classify the particles, strings and membranes arising from wrapped $p$--branes which have a charges conserved modulo some integer number $q$, and discuss  the catalyses of their annihilation by fluxes and ${\bb Z}_q$ gauge symmetry  associated
with those.
The actions of the type considered in this paper can be used to describe the effective field theory of such compactifications with networks of Dp-branes.


The rest of this paper is structured as follows. In sec. 2 we present the interacting action of the dynamical  system of open supermembrane, supergravity and a single three form matter multiplet which is the master system used later as a basis to construct the actions for more complicated interacting systems. The closed supermembrane action and its $\kappa$--symmetry is described in sec. 2.1, the interaction of dynamical supergravity with closed supermembrane is the subject of sec. 2.2.

In sec. 2.3 we discuss the breaking of the $\kappa$--symmetry in the case of open supermembrane and show that it can be partially restored by adding to the  open supermembrane action a certain  boundary term which can be interpreted as an action for closed superstring at the end of open supermembrane. The additional projection conditions on the $\kappa$--symmetry parameter at the worldsheet  presented there  suggest  that the open supermembrane (and supermembrane junctions) can preserve not more than one quarter of the spacetime supersymmetry. The spontaneous breaking of the three form gauge symmetry by superstring at the boundary of supermembrane and gauge fixed form of the boundary superstring action is also discussed in sec. 2.3. This section is finished by describing the most general form of the interacting action for the dynamical system under consideration, which includes, in particular, the mass term for the single 3-form multiplet.

The interacting system of open supermembrane, double three form supermultiplet and supergravity is described in sec. 3. In sec. 4 we present an  action for quite generic interacting system including, besides open supermembrane with superstring at its ends and double 3-form supergravity, a nonlinear interacting system of $n$ double-three form multiplets. This system is constructed with the use of special geometry and possesses symplectic $Sp(2n+2|{\bb Z})$ invariance provided the $2(n+1)$ charges carried by the open supermembrane transform as symplectic vector.
This interacting action, generalizing the action for the system with  closed supermembrane studied in \cite{Bandos:2018gjp}, can be used to investigate the role of open branes and branes at the boundary of open branes in the effective actions of the models originating in string compactifications. We conclude in sec. 5. Some useful equations  are collected in the Appendices.

\section{Open supermembrane  interacting with  single 3--forms matter and supergravity.}
 \label{sec2}

\subsection{Supermembrane action in the background of 3-form supergravity and  3-form matter}\label{3fsugra}


The action for a supermembrane in a supergravity background and also in the background of supergravity and  3-form matter multiplet(s) can be written in the following form
  \begin{eqnarray}\label{Sp=2:=}
  S_{p=2}= \int\limits_{W^3} d^3 \xi \sqrt{|h|}\,|{\cal{Z}}| +\int\limits_{W^3} {C}_3 \; . \qquad
\end{eqnarray}
In the first, Dirac-Nambu-Goto term of this action
$\xi^m=(\xi^0,\xi^1,\xi^2)$ are local coordinates on the worldvolume $W^3$ of the supermembrane, which is defined as a surface in superspace $\Sigma^{(4|4)}$
with coordinates $z^M=(x^\mu, \theta^{\check{\alpha}}, \bar{\theta}{}^{\check{\dot\alpha}})$
with the use of coordinate functions $z^M(\xi)=(x^\mu(\xi), \theta^{\check{\alpha}}(\xi), \bar{\theta}{}^{\check{\dot\alpha}}(\xi))$,
\begin{eqnarray}\label{W3in}
W^3\in \Sigma^{(4|4)}\; :\qquad z^M= z^M(\xi) \; .
\end{eqnarray}
$h=\det h_{mn}$ is the determinant of the induced metric
 \begin{eqnarray}\label{hmn=}
 h_{mn}= E_m^a \eta_{ab} E_n^b \; , \qquad
 E_m^a=\partial_m z^M(\xi) E_M^a (z(\xi))\; \qquad
\end{eqnarray}
which is constructed from the pull-back $E^a(z(\xi))= d\xi^m  E_m^a $ of the bosonic supervielbein
of the supergravity superspace,
 \begin{eqnarray}\label{EA=}
 E^A(z)=(E^a, E^\alpha ,\bar{E}{}^{\dot{\alpha}}) =dz^M E_M^A(z) \; . \qquad
\end{eqnarray}
Finally, ${\cal{Z}}$  denotes the pull-back ${\cal{Z}}({z}(\xi))$ of a  covariantly chiral superfield ${\cal Z}(z)$ of a special type which we describe below. Now we just notice that, as any covariantly chiral superfield, ${\cal Z}(z)$ obeys the  constraint
   \begin{eqnarray}\label{bDT2=0}
 \bar{{\cal D}}_{\dot\alpha} {\cal Z} = 0
\; ,   \qquad
\end{eqnarray}
where $\bar{{\cal D}}_{\dot\alpha}=- ({\cal D}_{\alpha})^*$ is the spinor covariant derivative defined by decomposition of the covariant differential on supervielbein,
\begin{eqnarray}\label{calD=}
{\cal D}=E^A {\cal D}_A = E^a {\cal D}_a + E^\alpha {\cal D}_\alpha +
\bar{E}{}^{\dot\alpha} \bar{{\cal D}}_{\dot\alpha}
\; .   \qquad
\end{eqnarray}
Supervielbein \eqref{EA=} is restricted by minimal supergravity constraints which we present in the Appendix A (see also \cite{Bandos:2002bx} and refs therein).

In the second, Wess-Zumino term of the supermembrane action \eqref{Sp=2:=}, $C_3$ is the pull-back of a 3-form potential defined in curved superspace and having the field strength 4-form expressed in terms of the above chiral superfield $ {\cal Z}$ by\footnote{To our best knowledge, the  closed 4-form \eqref{H4=Z} in supergravity superspace was first presented in \cite{Binetruy:1996xw} and its super-Weyl invariance was noticed in \cite{Kuzenko:2017vil}. }
\begin{eqnarray}\label{H4=Z}
{{H}}_{4}&=& dC_3= \frac 1 2 E^b\wedge E^a \wedge E^\alpha \wedge E^\beta \sigma_{ab\;\alpha\beta} \bar{ {\cal Z}} + \frac 1 2 E^b\wedge E^a \wedge E^{\dot\alpha} \wedge E^{\dot\beta} \tilde{\sigma}_{ab\;\dot\alpha\dot\beta} {\cal Z} + \qquad \nonumber \\ && + \frac 1 {12} E^c\wedge E^b\wedge E^a \wedge \epsilon_{abcd} E^\alpha \sigma^d_{\alpha\dot{\beta}} \bar{{\cal D}}{}^{\dot{\beta}}\bar{ {\cal Z}}  + \frac 1 {12} E^c\wedge E^b\wedge E^a \wedge \epsilon_{abcd} E^{\dot{\beta}} \sigma^d_{\alpha\dot{\beta}} {\cal D}{}^{\alpha}{\cal Z}  + \qquad \nonumber \\ && +\frac i {192} E^d\wedge E^c\wedge E^b\wedge E^a  \epsilon_{abcd}\left(( \bar{{\cal D}} \bar{{\cal D}}- 3R) \bar{ {\cal Z}} - ( {\cal D} {\cal D}- 3\bar{R}) {\cal Z}\right) \; .
\end{eqnarray}
Here $R=(\bar{R})^*$ and $G_a=(G_a)^*$ are main superfields of minimal (and variant) off-shell supergravity (see Appendix A).

The form ${{H}}_{4}$ is closed, $dH_4=0$,  when the supervielbein obeys the minimal supergravity constraints. However, the  requirement that it is exact, i.e. that there exists a 3-form $C_3$ such that $H_4=dC_3$, requires
the  chiral superfield $ {\cal Z}$ to be special, namely to be constructed in terms of real superfield prepotential ${\cal P}={\cal P}^*$,
 \begin{eqnarray}\label{T2=bDbDP}
  {\cal Z}= -\frac 1 4 \left(\bar{{\cal D}}_{\dot\alpha}\bar{{\cal D}}{}^{\dot\alpha}- R\right) {\cal P}
\; . \qquad
\end{eqnarray}
As a result, the component content of ${\cal Z}$ is different from that of the usual chiral superfield
$\Phi=  -\frac 1 4 \left(\bar{{\cal D}}_{\dot\alpha}\bar{{\cal D}}{}^{\dot\alpha}- R\right) {\bb K}$  constructed from the complex potential $ {\bb K}\not= {\bb K}^*$: the $F$-component of that superfield is given by a complex linear combination of real scalar and a divergence of a real vector instead of two real scalars (scalar and pseudoscalar) in the case of $\Phi$ (see e.g. \cite{Bandos:2019qok} and refs. therein for more details). Hence the name of single three form supermultiplet for the field content of the special chiral  superfield \eqref{T2=bDbDP} with an arbitrary superfield  ${\cal P}$.

Now, the real 3-form potential $C_{3}$, the pull--back of which  enters  the second term in (\ref{Sp=2:=}), is expressed  in terms of the same real superfield ${\cal P}$ by
\begin{eqnarray}
\label{C3=cP}   C_{3}  &=&  -iE^a \wedge E^\alpha \wedge \bar E^{\dot\alpha}  \sigma_{a\alpha\dot\alpha} {\cal P} - {1\over 4}  E^b\wedge E^a \wedge  E^\alpha
\sigma_{ab\; \alpha}{}^{\beta}{\cal D}_{\beta}{\cal P} +\qquad \nonumber \\ &&+ {1\over 4}  E^b\wedge E^a \wedge  \bar E^{\dot\alpha}
\tilde{\sigma}_{ab}{}^{\dot\beta}{}_{\dot\alpha}
\bar{{\cal D}}_{\dot\beta}{\cal P}
 + {1\over 48}  E^c \wedge E^b \wedge E^a  \epsilon_{abcd}\left(\tilde{\sigma}^{d\dot{\alpha}\alpha}[{\cal D}_\alpha ,\bar{{\cal D}}_{\dot\alpha}]{\cal P}+2G^d{\cal P}
\right)
 \, .    \qquad
\end{eqnarray}

Of course, \eqref{C3=cP} is the gauge fixed  form of the potential corresponding to the field strength \eqref{H4=Z}. However, there exists a residual gauge invariance with respect to additive transformations of real prepotential superfield ${\cal P}$ with real linear superfield ${\bb L}$,
\begin{eqnarray} \label{vcP=tL}
\delta {\cal P}={\bb L} \, , && \qquad \\
\label{DDbbL=0=}
&& \left(\bar{{\cal D}}_{\dot\alpha}\bar{{\cal D}}{}^{\dot\alpha}- R\right){\bb L}=0\; , \qquad
\left({{\cal D}}^{\alpha}{{\cal D}}{}_{\alpha}- \bar{R}\right){\bb L}=0   \; .
\end{eqnarray}
Such transformation of the prepotential results in the gauge transformation of the superspace 3-form \eqref{C3=cP}
\begin{eqnarray} \label{vC3=dal2}
\delta {C}_3=d\alpha_2 \;  \qquad
\end{eqnarray}
by closed 3--form $d\alpha_2$ constructed from the  real linear superfield  ${\bb L}$ \eqref{DDbbL=0=} as follows
\begin{eqnarray} \label{dal2=}    d\alpha_2&=& -i E^a \wedge E^\alpha \wedge \bar E^{\dot\alpha}  \sigma_{a\alpha\dot\alpha} {\bb L} - {1\over 4}  E^b\wedge E^a \wedge  E^\alpha
\sigma_{ab\; \alpha}{}^{\beta}{\cal D}_{\beta}{\bb L} +\qquad \nonumber \\ &&+ {1\over 4}  E^b\wedge E^a \wedge  \bar E^{\dot\alpha}
\tilde{\sigma}_{ab}{}^{\dot\beta}{}_{\dot\alpha}
\bar{{\cal D}}_{\dot\beta}{\bb L}
 + {1\over 48}  E^c \wedge E^b \wedge E^a  \epsilon_{abcd}\left(\tilde{\sigma}^{d\dot{\alpha}\alpha}[{\cal D}_\alpha ,\bar{{\cal D}}_{\dot\alpha}]{\bb L}+2G^d{\bb L}
\right)
\, . \qquad
\end{eqnarray}
Clearly, $d\alpha_2={C}_3\vert_{{\cal P}\mapsto {\bb L}}$.

The {\it closed} supermembrane action is invariant under local fermionic $\kappa$--symmetry transformations of the coordinate functions
 \begin{eqnarray}\label{ikappaEA=}
 i_\kappa E^a :=\delta_\kappa z^M E_M^a =0 \; , \qquad
 i_\kappa E^\alpha :=\delta_\kappa z^M E_M^\alpha =\kappa^\alpha \; , \qquad
  i_\kappa E^{\dot\alpha} :=\delta_\kappa z^M E_M^{\dot\alpha}  =\bar{\kappa}{}^{\dot\alpha} \;  \qquad
 \end{eqnarray}
 the fermionic parameters of which obey the conditions
 \begin{eqnarray}\label{kappa=Gbk}
 \kappa_\alpha = - i \frac {{\cal Z}}  {|{\cal Z}|} \; \Gamma_{\alpha\dot\alpha}\bar{\kappa}{}^{\dot\alpha} \; , \qquad
 \bar{\kappa}{}_{\dot\alpha}= - i \frac {\bar{\cal Z}}  {|{\cal Z}|} \;  \kappa^\alpha \Gamma_{\alpha\dot\alpha} \; , \qquad
 \end{eqnarray}
where
 \begin{eqnarray}\label{G=}
\Gamma_{\alpha\dot\alpha}=\frac i {3!\sqrt{h}}\; \sigma^a_{\alpha\dot\alpha} \epsilon_{abcd} \epsilon^{mnk} E_m^b E_n^cE_k^d \; ,\qquad
 \end{eqnarray}
is imaginary, $(\Gamma_{\alpha\dot\alpha})^*=- \Gamma_{\alpha\dot\alpha}$, and
obeys $\Gamma_{\alpha\dot\alpha}\Gamma^{\dot\alpha\beta }=\delta_{\alpha}{}^{\beta }$.

Finally let us comment on the dimension of special chiral superfield ${\cal Z}$. If we read Eq. \eqref{Sp=2:=} literally, we should conclude that the dimension of  ${\cal Z}$ is 3 in the mass units, $[{\cal Z}]=M^{3}$. This is because the tension of the supermembrane,  $T_2$, is included in ${\cal Z}$ as a multiplier. To make its presence explicit we should redefine
${\cal Z}\mapsto T_2{\cal Z}$ and consider ${\cal Z}$ to be  dimensionless,  $[{\cal Z}]=M^{0}$.
We can also consider  \eqref{Sp=2:=} as an action with $T_2$ formally set to be 1 and containing a dimensionless ${\cal Z}$. We will prefer such interpretation of our action.

\subsection{Supergravity interacting with closed supermembrane}

As we have already stated, the action (\ref{Sp=2:=}) can describe the supermembrane moving in the background of a three-form supergravity as well as in the background of supergravity and 3--form matter  multiplet(s).
In the first case the above special chiral superfield ${\cal Z}$ should be treated as conformal compensator of a 3-form supergravity.

The name of 3--form supergravity is attributed to two  variant formulations of minimal supergravity \cite{Gates:1980ay,Gates:1980az}, presently referred to as single three form supergravity and double three form supergravity \cite{Farakos:2017jme}.
 In the (super--)Weyl invariant formulation of these versions of ${\cal N}=1$ supergravity  the conformal compensator of minimal supergravity ${\cal Z}$ has a  special form: it is  expressed in terms of real prepotential superfield ${\cal P}$ as in (\ref{T2=bDbDP}). In the case when this prepotential is an independent ('fundamental') superfield, we arrive at single three form supergravity \cite{Gates:1980az,Ovrut:1997ur,Buchbinder:1988tj,Kuzenko:2005wh,Bandos:2011fw,Bandos:2012gz,Farakos:2016hly}
in its Weyl invariant formulation of \cite{Kuzenko:2005wh}.
If the chiral compensator is expressed in terms of composite real prepotential given by real or imaginary part of a complex linear superfield $\Sigma$,
\begin{eqnarray}\label{cP=ReSi}
{\cal P}=\Im{\rm m} \Sigma :=
\frac i {2} (\bar{\Sigma}-\Sigma )
\; , \qquad
({\cal D}^\alpha {\cal D}_\alpha -\bar{R})\Sigma=0\; , \qquad
(\bar{{\cal D}}_{\dot\alpha}\bar{{\cal D}}{}^{\dot\alpha} -R)\bar{\Sigma}=0\; , \qquad
\end{eqnarray}
we are dealing with double three form supergravity, which was actually described already in
\cite{Stelle:1978ye} and \cite{Ogievetsky:1980qp}. Its coupling to matter and application to the effective theory of string compactifications was the subject of recent \cite{Farakos:2017jme}.

The action for the interacting system of double three form supergravity and closed supermembrane was presented in
\cite{Kuzenko:2017vil}. The dynamical system including also a set of nonlinearly self interacting double 3--form matter multiplets was described and studied in \cite{Bandos:2018gjp}. The single three form supergravity interacting with supermembrane was the subject of  \cite{Ovrut:1997ur,Bandos:2011fw,Bandos:2012gz}. Here we will  consider the case of supergravity interacting with  open supermembrane and superstring at the boundary of the open supermembrane.

But first let us write the action for the interacting system of supergravity and closed membrane in super-Weyl invariant formulation of supergravity. It reads
\begin{eqnarray}\label{S=Ssg+Sp2}
S=S_{sugra}+S_{p=2}\; , \qquad
\end{eqnarray}
where $S_{p=2}$ has the form of (\ref{Sp=2:=}), with (\ref{T2=bDbDP}) and  (\ref{C3=cP}), and\footnote{As super-Weyl invariant action for three-form supergravity Eq. \eqref{Ssg=} was discussed in \cite{Kuzenko:2005wh}. See \cite{Buchbinder:1995uq}
for the description of new minimal and nonminimal off-shell formulations of supergravity as super-Weyl-invariant couplings of the old minimal supergravity to a compensating supermultiplet.}
\begin{eqnarray}\label{Ssg=}
S_{sugra}=-\frac 3 {4\kappa^2} \int d^8z \, E\, ({\cal Z}\bar{{\cal Z}})^{\frac 1 3 } -\frac m {2\kappa^2} \left( \int d^6\zeta_L \, {\cal E}\, {\cal Z} + c.c.\right)
\; . \qquad
\end{eqnarray}
Here $E=sdet (E_M^A(z))$ is the superdeterminant (Berezenian) of the supervielbein,
$m$ is the gravitino mass, proportional to the cosmological constant ($m=0$ for the case of Poincar\'e supergravity)
and $d^6\zeta_L \, {\cal E}\,$ is the chiral measure (see \cite{Wess:1992cp} and refs. therein). This  is related to the complete superspace measure $d^8z E$ by
\begin{eqnarray}\label{d8z=d6zL}
 \int d^8z \, E\, {\bb Y} = -\frac 1 2 \int d^6\zeta_L \, {\cal E}\, \left(\bar{\cal D}\bar{\cal D} - R\right){\bb Y}
\; , \qquad
\end{eqnarray}
where ${\bb Y}$ is an arbitrary superfield.

The super-Weyl transformations  leaving invariant  the action \eqref{S=Ssg+Sp2}, \eqref{Ssg=}, \eqref{Sp=2:=}
are described in Appendix B (see Eqs. \eqref{supW=m4Db}--\eqref{supW=cZ}and \eqref{bDUp=0}). This can be used to set the chiral superfield ${\cal Z}$ equal to unity.
The super-Weyl symmetry of the action \eqref{Ssg=} is thus realized by St\"uckelberg mechanism with a pure gauge superfield  ${\cal Z}$. Hence the name of   conformal compensator used for ${\cal Z}$ superfield in the action \eqref{Ssg=}.

\subsection{Interaction with supergravity and single 3-form matter multiplet}

The case of simplest interacting system of supergravity, supermembrane and a 3-form matter multiplet is described by the action
\begin{eqnarray}\label{S=Ssg+m+Sp2}
S=S_{sugra+matter}+S_{p=2}\; , \qquad
\end{eqnarray}
where $S_{p=2}$ has the form of (\ref{Sp=2:=}) and
\begin{eqnarray}\label{Ssg+1=}
S_{sugra+matter}=-\frac 3 {4\kappa^2} \int d^8z \, E\, \Omega({\cal Z},\bar{{\cal Z}}) -\frac 1 {2\kappa^2} \left( \int d^6\zeta_L \, {\cal E}\, {\cal W}({\cal Z}) + c.c.\right)
\; . \qquad
\end{eqnarray}
Here ${\cal W}({\cal Z}) $ is superpotential,
\begin{eqnarray}\label{Om=e-13K}
 \Omega({\cal Z},\bar{{\cal Z}})= e^{- \frac {\kappa^2} 3 K({\cal Z},\bar{{\cal Z}})}
\; , \qquad
\end{eqnarray}
and $K({\cal Z},\bar{{\cal Z}})$ is the K\"ahler potential. The action \eqref{Ssg+1=} is invariant under the super-Weyl transformations
\eqref{supW=m4Db}, \eqref{supW=m4Df}, \eqref{supW=m4Dbf} with $\Upsilon=\Upsilon({\cal Z})$, $\bar\Upsilon=\bar\Upsilon(\bar{\cal Z})$,   supplemented by the K\"ahler trasformations of the K\"ahler potential,
$$
 K({\cal Z},\bar{{\cal Z}}) \mapsto  K({\cal Z},\bar{{\cal Z}}) + 6\Upsilon({\cal Z}) + 6\bar\Upsilon(\bar{\cal Z})
$$
and
$
{\cal W}({\cal Z})\mapsto  {\cal W}({\cal Z})e^{-6\Upsilon({\cal Z})}
$. In the case of nonvanishing superpotential, these transformations  can be used to gauge this to a constant $m$,
\begin{eqnarray}\label{K->K+lnW}
 K({\cal Z},\bar{{\cal Z}}) \mapsto {\cal  K}({\cal Z},\bar{{\cal Z}}) =K({\cal Z},\bar{{\cal Z}}) + \frac 2 {\kappa^2}\ln |{\cal W}({\cal Z})| - \frac 2 {\kappa^2} \ln |m| \; , \qquad {\cal W}({\cal Z})\mapsto  m \; . \qquad
\end{eqnarray}
In the case of ${\cal W}({\cal Z})=m{\cal Z}$ and $K({\cal Z},\bar{{\cal Z}})= -\frac 2 {\kappa^2} \ln |{\cal Z}|$, in which \eqref{Ssg+1=} reduces to \eqref{Ssg=}, the transformation \eqref{K->K+lnW} removes the chiral superfield ${\cal Z}$ from the action. This indicates that, as stated, the action \eqref{Ssg=} describes the (3-form) supergravity only.

\subsection{Superstring at the boundary  of open  supermembrane coupled to supergravity and single 3-form matter multiplet}

When the supermembrane worldvolume is not closed, $\partial W^3= W^2\not= \bigcirc\!\!\!\!\!\big/\;$,
 its action (\ref{Sp=2:=}) is not invariant under the above described $ \kappa$-symmetry,
\begin{eqnarray}\label{vkSp2=open}
\delta_\kappa S_{p=2} &=& \int\limits_{W^2=\partial W^3} \; i_\kappa C_3\; ,  \qquad \\
\label{ikC3=}
 i_\kappa C_{3}  &=&  -iE^a \wedge E^\alpha\sigma_{a\alpha\dot\alpha} \bar{\kappa}{}^{\dot\alpha} {\cal P}
 -iE^a \wedge \bar E^{\dot\alpha} \, {\kappa}{}^{\alpha} \sigma_{a\alpha\dot\alpha}  {\cal P} - \qquad\nonumber \\
 && - {1\over 4}  E^b\wedge E^a \left(\kappa^\alpha
\sigma_{ab\; \alpha}{}^{\beta}{\cal D}_{\beta}{\cal P} -
\tilde{\sigma}_{ab}{}^{\dot\beta}{}_{\dot\alpha} \bar{\kappa}{}^{\dot\alpha}\,
\bar{{\cal D}}_{\dot\beta}{\cal P}\right)
 \, .    \qquad
\end{eqnarray}
Neither  the open supermembrane action (\ref{Sp=2:=}) is invariant under the gauge transformations \eqref{vcP=tL}: we find
\eqref{vC3=dal2} and
\begin{eqnarray}\label{vkSp2=open}
\delta_{gauge} S_{p=2} &=& \int\limits_{W^2=\partial W^3} \;   \alpha_2\; ,
\end{eqnarray}
where $\alpha_2$ is defined by \eqref{dal2=}.

To compensate these nonvanishing variations, it is necessary to put at the boundary of supermembrane a superstring.
For the gauge symmetry the  mechanism of compensation refers to the Wess--Zumino term of the superstring action
which is given by integral over the worldsheet of a 2-form potential $B_2$,
\begin{eqnarray} \label{intB2=intH3}
 -\int\limits_{W^2} B_2=  - \int\limits_{W^3} d{B}_2 \equiv - \int\limits_{W^3} {H}_3
\, . \qquad
\end{eqnarray}
The sum of the Wess-Zumino terms of string and membrane
\begin{eqnarray} \label{mem+str=WZ}
\int\limits_{W^3} {C}_3- \int\limits_{W^2} {B}_2=  \int\limits_{W^3} ( {C}_3- d{B}_2)
\,  \qquad
\end{eqnarray}
will be invariant under 3-form gauge transformations \eqref{vC3=dal2}, \eqref{dal2=} if 2-form potential  transforms under these as a St\"{u}ckelberg field,
\begin{eqnarray} \label{vB2=al2}
\delta {C}_3=d\alpha_2 \; , \qquad \delta {B}_2=\alpha_2
\,  . \qquad
\end{eqnarray}

This is possible if $B_2$ in \eqref{intB2=intH3} is the pull-back of the superspace 2-form with the field strength expressed by
\begin{eqnarray} \label{H3=dB2=}   H_{3}  &=&  dB_2= -i E^a \wedge E^\alpha \wedge \bar E^{\dot\alpha}  \sigma_{a\alpha\dot\alpha} L - {1\over 4}  E^b\wedge E^a \wedge  E^\alpha
\sigma_{ab\; \alpha}{}^{\beta}{\cal D}_{\beta}L +\qquad \nonumber \\ &&+ {1\over 4}  E^b\wedge E^a \wedge  \bar E^{\dot\alpha}
\tilde{\sigma}_{ab}{}^{\dot\beta}{}_{\dot\alpha}
\bar{{\cal D}}_{\dot\beta}L
 + {1\over 48}  E^c \wedge E^b \wedge E^a \epsilon_{abcd}\left(\tilde{\sigma}^{d\dot{\alpha}\alpha}[{\cal D}_\alpha ,\bar{{\cal D}}_{\dot\alpha}]L+2G^dL
\right)
 \,   \qquad
\end{eqnarray}
($H_3=C_3\vert_{{\cal P}\mapsto L}$)
in terms of  the real tensor multiplet $L$
\begin{eqnarray}\label{DDL=0=}
\left(\bar{{\cal D}}_{\dot\alpha}\bar{{\cal D}}{}^{\dot\alpha}- R\right)L=0\; , \qquad
\left({{\cal D}}^{\alpha}{{\cal D}}{}_{\alpha}- \bar{R}\right)L=0   \;
\end{eqnarray}
which is  transformed  as St\"uckelberg superfield under \eqref{vcP=tL},
\begin{eqnarray} \label{vcP=tL+}
\delta {\cal P}={\bb L} \, , \qquad \delta L ={\bb L} \, . \qquad
\end{eqnarray}

In the absence of open supermembrane, the action of closed superstring with Wess--Zumino term \eqref{intB2=intH3}
contains also a Nambu-Goto term including the pull--back to worldline of the real linear superfield {{L}},
\begin{eqnarray} \label{SNG=L0}
 {1\over 2} \int\limits_{W^2}d^2\sigma\sqrt{-\gamma} \, |{{L}}|\; .
\end{eqnarray}
Here $d^2\sigma=d\sigma^0\wedge d\sigma^1$, $\sigma^i =(\sigma^0, \sigma^1)$ are local worldsheet coordinates and
$\gamma=\det \gamma_{ij}$ is the determinant of the metric induced on the worldsheet $W^2$,
\begin{eqnarray} \label{gij=}
\gamma_{ij}= E^a_i\eta_{ab} E^b_j \; , \qquad  E^a_i=\partial_i z^M(\sigma) E_M^a (z(\sigma))
\; .
\end{eqnarray}

When the string is situated at the boundary of membrane, the term \eqref{SNG=L0} should be modified to $ {1\over 2} \int\limits_{W^2}d^2\sigma\sqrt{-\gamma} \, |{\cal P}- {{L}}|$ as, after such a modification, the Nambu-Goto term  will respect the gauge symmetry \eqref{vcP=tL+} which also leaves invariant the sum of the Wess--Zumino terms of the superstring and the supermembrane as well as the
Dirac-Nambu-Goto term of the supermembrane action. Thus we arrive at the following action for superstring at the boundary of supermembrane
 \begin{eqnarray}\label{Sp=1:=}
  S_{p=1} &=& {1\over 2} \int\limits_{W^2}d^2\sigma\sqrt{-\gamma} \, |{\cal P}- {L}| - \int\limits_{W^2} B_2 , \;  \qquad
\end{eqnarray}
where the field strength of $B_2$ has the form of \eqref{H3=dB2=}, $L$ is the pull--back of a real linear superfield obeying \eqref{DDL=0=}. Finally
${\cal P}$ in \eqref{Sp=1:=} is the pull--back to the worldsheet of the real prepotential superfield defining the special chiral superfield through Eq. \eqref{T2=bDbDP} and the three form potential through Eq. \eqref{C3=cP}.

The  actions given by the sum of \eqref{Sp=1:=} and  \eqref{Sp=2:=} is also invariant under the local fermionic $\kappa$--symmetry
\eqref{ikappaEA=} with parameters restricted, besides \eqref{kappa=Gbk}, by the projection conditions
\begin{eqnarray}\label{kappa-str}
\kappa_\alpha = \frac {{\cal P}-{\bb L}}{|{\cal P}-{\bb L}|} \,  P_\alpha{}^\beta \kappa_\beta\, , \qquad
\bar\kappa_{\dot\alpha} = \, \frac {{\cal P}-{\bb L}}{|{\cal P}-{\bb L}|} \,  \bar P_{\dot\alpha}{}^{\dot\beta}\bar \kappa_{\dot\beta}\,
\end{eqnarray}
where
\begin{eqnarray}\label{proj-str}
P_\beta{}^\alpha
&=& \frac 1 {2\sqrt{-\gamma}}\epsilon^{ij} E_i^a E_j^b \sigma_{ab}{}_\beta{}^\alpha, \qquad
\bar P_{\dot\alpha}{}^{\dot\beta}= (P_{\alpha}{}^{\beta})^* =  - \frac 1 {2\sqrt{-\gamma}}\epsilon^{ij} E_i^a E_j^b \tilde{\sigma}_{ab}{} ^{\dot\beta}{}_{\dot\alpha}\;\end{eqnarray}
obey
\begin{eqnarray}\label{proj-str}
P^2&=&{\mathbb I}\; , \qquad \bar{P}{}^2={\mathbb I}\; .
\end{eqnarray}
For a particular case of superstring at the boundary of open supermembrane interacting with Veneziano-Yankelovich effective description of the SYM theory the (flat superspace version of the) action \eqref{Sp=1:=} was found in \cite{Bandos:2019qok} where the above $\kappa$--symmetry was also presented.

The new property of the interacting system including supergravity, which follows from its the diffeomorphism gauge invariance, is that the superstring and supermembrane Goldstone fields, this is to say bosonic and fermionic coordinate functions, become St\"uckelberg (pure gauge) fields which do not carry degrees of freedom. This allows to fix their values by imposing, e.g.
\begin{eqnarray}\label{x=sigma}
x^i(\sigma)= \sigma^i \; , \qquad x^2(\sigma)= 0  \; , \qquad x^3(\sigma)= 0  \; , \qquad \theta^{\check{\alpha}}(\sigma)= 0
\; , \qquad \bar{\theta}{}^{\check{\dot\alpha}}(\sigma)= 0
\end{eqnarray}
in the case of superstring and
\begin{eqnarray}\label{x=xi}
x^m(\xi)= \xi^m \; , \qquad   x^3(\xi)= 0  \; , \qquad \theta^{\check{\alpha}}(\xi)= 0
\; , \qquad \bar{\theta}{}^{\check{\dot{\alpha}}}(\xi)= 0
\end{eqnarray}
in the case of supermembrane.

Let us stress that, when dynamical supergravity described by the superfields which are varied in the action, is not present, like in flat superspace system discussed in \cite{Bandos:2019qok}, Eqs. \eqref{x=xi} and \eqref{x=sigma}  describe a particular configuration of open supermembrane and  superstring at the boundary of this supermembrane. In contrast, when supergravity is dynamical the diffeomorphism invariance is the gauge symmetry of the system and \eqref{x=xi} and  \eqref{x=sigma} describe just gauge fixed conditions for such a symmetry spontaneously broken by open supermembrane and superstring at its boundary.

The superstring at the boundary of supermembrane also breaks the gauge symmetry (\ref{vcP=tL}), (\ref{vC3=dal2}), (\ref{dal2=}), characteristic for the three form potential. When action is written with the use of the St\"uckelberg real linear superfield $L$, as in \eqref{Sp=1:=}, this symmetry is formally maintained (realized dynamically) as ${\cal P}-L$ is invariant under (\ref{vcP=tL+}). However, we can fix the gauge under this symmetry by setting $L=0$ and in this gauge \eqref{Sp=1:=} reduces to
 \begin{eqnarray}\label{Sp=1=cP}
  S_{p=1}\vert_{_{L=0}} &=& {1\over 2} \int\limits_{W^2}d^2\sigma\sqrt{-\gamma} \, |{\cal P}|  .  \;  \qquad
\end{eqnarray}
Notice that the Wess--Zumino term of the superstring vanishes in this gauge. The remaining Nambu-Goto type term
\eqref{Sp=1=cP} is sufficient to compensate \eqref{vkSp2=open} with \eqref{ikC3=} provided the
$\kappa$-symmetry parameter is restricted, besides \eqref{kappa=Gbk}, also by the condition ({\it cf.} \eqref{kappa-str})
\begin{eqnarray}\label{kappa-strP}
\kappa_\alpha = \frac {{\cal P}}{|{\cal P}|} \,  P_\alpha{}^\beta \kappa_\beta\, , \qquad
\bar\kappa_{\dot\alpha} = \, \frac {{\cal P}}{|{\cal P}|} \,  \bar P_{\dot\alpha}{}^{\dot\beta}\bar \kappa_{\dot\beta}\, ,
\end{eqnarray}
where $ P_\alpha{}^\beta = (\bar{P}_{\dot\alpha}{}^{\dot\beta})^*$ is defined in \eqref{proj-str}.

When restoring the membrane tension in the gauge fixed action \eqref{Sp=1=cP}, it takes the form
$S_{p=1}\vert_{_{L=0}} = {T_2\over 2} \int\limits_{W^2}d^2\sigma\sqrt{-\gamma} \, |{\cal P}|$ which makes manifest that the effective tension  of the string at the boundary of supermembrane is defined  by the supermembrane tension:  $T_1(\sigma)= T_2 |{\cal P}(z(\sigma)|$.
To stress this, it is instructive to write once the action of open supermembrane and the superstring at its boundary with the membrane tension written explicitly:
  \begin{eqnarray}\label{Sp=2+1=T}
  S_{p=2}+S_{p=1}= T_2 \int d^3 \xi \sqrt{|h|}\,|{\cal{Z}}| +T_2\int\limits_{W^3} ({C}_3-H_3) +
  {T_2\over 2} \int\limits_{W^2=\partial W^3}d^2\sigma\sqrt{-\gamma} \, |{\cal P}- {L}| \; . \qquad
\end{eqnarray}

Notice that in the $L=0$ gauge the form of the action looks like  a  counterpart of the Fayet--Iliopoulos term, but  with the real prepotential superfield $V$ of the $U(1)$ SYM model replaced by the real prepotential superfield
${\cal P}$ of the three form multiplet, and with superspace integration replaced by the integration over the worldsheet. This later results in the explicit breaking of the three-form gauge symmetry (\ref{vcP=tL}) in the (gauge fixed) action including \eqref{Sp=1=cP}, while when the action contains \eqref{Sp=1:=}, the three form gauge symmetry is maintained  but realized with St\"uckelberg mechanism as in  (\ref{vcP=tL+}).

As far as the  breaking of the three form gauge symmetry is allowed, we can add one more  term to the supergravity plus matter part \eqref{Ssg+1=} of the action \eqref{S=S+}. These is
the mass term for the 3-form matter multiplet, $\int d^8z \, E\, \frac {{\cal P}^2}{({\cal Z}\bar{{\cal Z}})^{1/3}}$ \cite{Farrar:1997fn}. As we have already introduced the St\"uckelberg real linear superfield $L$ in the bulk, thus stressing the spontaneous character of the breaking of
the 3-form gauge symmetry by superstring at the boundary of supermembrane, we can write this term with maintaining formal gauge invariance as
$\int d^8z \, E\, \frac {({\cal P}-L)^2}{({\cal Z}\bar{{\cal Z}})^{1/3}}$.
Then the most general (up to inclusion of higher derivative terms) $S_{sugra+matter}$ part of the action \eqref{S=S+} reads
\begin{eqnarray}\label{Ssg+1+=}
S_{sugra+matter}&=&-\frac 3 {4\kappa^2} \int d^8z \, E\, \Omega({\cal Z},\bar{{\cal Z}})
 -\frac 1 {2\kappa^2} \left( \int d^6\zeta_L \, {\cal E}\, {\cal W}({\cal Z}) + c.c.\right) - \qquad \nonumber \\ && - {\frak {m}}^4 \int d^8z \, E\, \frac {({\cal P}-L)^2}{({\cal Z}\bar{{\cal Z}})^{1/3}}
\; . \qquad
\end{eqnarray}
Here ${\frak {m}}$ is a constant of dimension of  mass. The remaining parts of the interacting action
\eqref{S=S+} are given in Eq. \eqref{Sp=2+1=T}. (This latter clearly indicates that the mass dimension of ${\cal Z}$ and ${\cal P}$ in \eqref{Ssg+1+=} is 0  and -1, respectively).

One might observe the possibility to add to the action a true counterpart of the Fayet--Iliopoulos term constructed from the real prepotential superfield: $\int d^8z \, E\,  {\cal P}$. However, taking into account the relation of the chiral and full superspace integration measure \eqref{d8z=d6zL}  it is easy to observe that actually this is an equivalent form of the F-term with linear superpotential for the special chiral superfield ${\cal Z}$ \eqref{T2=bDbDP}, $\int d^8z \, E\,  {\cal P}= \int d^6\zeta_L \, {\cal E}\,  {\cal Z}+c.c.$.

Notice that the form of the mass term in \eqref{Ssg+1+=} is fixed by the requirement of the super--Weyl invariance under (\ref{supW=m4Db})--\eqref{supW=cZ} with  \eqref{bDUp=0}. For the case of Veneziano-Yankelovich effective theory of SYM such a term was considered in \cite{Farrar:1997fn} and \cite{Bandos:2019qok}.

\section{Interacting system of double three form multiplets, supergravity and open supermembrane}

The open supermembrane part of the action for the interacting systems including a number of different 3-form matter multiplets and supergravity can be easily obtained from the above action for the case of supergravity interacting with a single three form multiplet.
The key point is to define the composite special chiral superfield ${\cal Z}$ and its real prepotential superfield ${\cal P}$ in terms of several 'fundamental' superfields. No need to stress that such a redefinition generically would result in a possible changes of the superfield matter part of the action and in any case would produce a different set of the equations of motion.

Below we would like to discuss the actions and symmetry of such  interacting systems beginning from the case of open supermembrane coupled to the double three form matter and supergravity.

A particular case of composite special chiral superfield  ${\cal Z}$ is reached when the prepotential in (\ref{T2=bDbDP}) is constructed as
\begin{eqnarray}\label{cP=ImSi}
{\cal P}=\Im {\rm m}\Sigma :=
\frac i 2 (\bar{\Sigma}-\Sigma )
\; \qquad
\end{eqnarray}
from the complex linear superfield $\Sigma$ obeying
\begin{eqnarray}\label{DDSi=bRSi}
({\cal D}^\alpha {\cal D}_\alpha -\bar{R})\Sigma=0\; . \qquad
\end{eqnarray}
Eq. (\ref{DDSi=bRSi}) is solved by
\begin{eqnarray}\label{Si=DXi}
\Sigma={\cal D}_\alpha \Xi^\alpha \;  \qquad
\end{eqnarray}
with an independent spinor superfield $\Xi^\alpha$.

A special chiral superfield ${\cal Z}$ defined in (\ref{T2=bDbDP}) and (\ref{cP=ImSi}), which we denote below by $\frac i 2 {\cal S}$,
\begin{eqnarray}\label{S=Si}
{{\cal S}} = \frac 1 4 \left(\bar{{\cal D}}_{\dot\alpha}\bar{{\cal D}}{}^{\dot\alpha}- R\right)\Sigma
 \; , \qquad \bar{{\cal S}} = \frac 1 4 \left({{\cal D}}^{\alpha}{{\cal D}}{}_{\alpha}- \bar{R}\right)\bar{\Sigma}   \; ,
\end{eqnarray}
has as  its F-component a linear combination of two divergences of real vectors (instead of two scalars in the case of usual chiral superfield
$\Phi$, see \cite{Farakos:2017jme} and refs. therein for details). Hence the name of double three form multiplet for the component content of the special chiral superfield ${{\cal S}}$.

There is also a related superspace reason for such a name. With ${\cal Z}=\frac i 2 {\cal S}$ defined in (\ref{T2=bDbDP}) and (\ref{cP=ImSi}), the real exact form (\ref{H4=Z}) is equal to  doubled real part of the complex exact form
\begin{eqnarray}\label{H4=2ReH4L}
 H_4=dC_3 &=& 2\Re {\rm e} \bar{{\cal H}}_{4}\; , \qquad \\ \label{H4L=bcZ}
\bar{{\cal H}}_{4}= d\bar{{\cal A}}_3&=& -\frac i 4 E^b\wedge E^a \wedge E^\alpha \wedge E^\beta \sigma_{ab\;\alpha\beta} \bar{{\cal S}} - \qquad \nonumber \\ &&  - \frac i {4!} E^c\wedge E^b\wedge E^a \wedge \epsilon_{abcd} E^\alpha \sigma^d_{\alpha\dot{\beta}} \bar{{\cal D}}{}^{\dot{\beta}}\bar{{\cal S}}  + \qquad \nonumber \\ &&+\frac 1 {384} E^d\wedge E^c\wedge E^b\wedge E^a  \epsilon_{abcd}( \bar{{\cal D}} \bar{{\cal D}}- 3R) \bar{{\cal S}} \; .
\end{eqnarray}
The complex three form potential for \eqref{H4L=bcZ} can be chosen to be
\begin{eqnarray} \label{bcA3=bSi}   \bar{{\cal A}}{}_{3}  &=&   {1\over 2} E^a \wedge E^\alpha \wedge \bar E^{\dot\alpha}  \sigma_{a\alpha\dot\alpha} \overline{\Sigma} - {i\over 8}  E^b\wedge E^a \wedge  E^\alpha
\sigma_{ab\; \alpha}{}^{\beta}{\cal D}_{\beta}\overline{\Sigma} +\qquad \nonumber \\ &&+ {i\over 8}  E^b\wedge E^a \wedge  \bar E^{\dot\alpha}
\tilde{\sigma}_{ab}{}^{\dot\beta}{}_{\dot\alpha}
\bar{{\cal D}}_{\dot\beta}\overline{\Sigma}
 + {1\over 3!}  E^c \wedge E^b \wedge E^a  \bar{{\cal A}}{}_{abc}
 \,   \qquad
\end{eqnarray}
with
\begin{eqnarray} \label{bcAabc=bSi}
  \bar{{\cal A}}{}_{abc}&=\frac{i}{16}\epsilon_{abcd}\left(\tilde{\sigma}^{d\dot{\alpha}\alpha}{\cal D}_\alpha\bar{{\cal D}}_{\dot\alpha}\overline{\Sigma}-2i {\cal D}^d\overline{\Sigma}+G^d\overline{\Sigma}
\right)
\qquad \nonumber \\ &=\frac{i}{32}\epsilon_{abcd}\left(\tilde{\sigma}^{d\dot{\alpha}\alpha}[{\cal D}_\alpha ,\bar{{\cal D}}_{\dot\alpha}]\overline{\Sigma}+2G^d\overline{\Sigma}
\right)
\, . \qquad
\end{eqnarray}

The real 3--form potential for \eqref{H4=Z} is now given by (twice the) real part of the complex potential,
\begin{eqnarray}
\label{C3=2ReH4L}
C_3 &=& 2\Re {\rm e} \bar{{\cal A}}_{3} ={{\cal A}}_{3}+ \bar{{\cal A}}_{3}\; ,
\end{eqnarray}
and  we actually have two gauge invariances of the type (\ref{vC3=dal2}), a one--parametric combination of which should act on our St\"uckelberg two--form,
\begin{eqnarray}
 \label{vcA3=db2}
\delta {{\cal A}}_{3}=d\beta_2 \; , \qquad
\delta  \bar{{\cal A}}_{3} =d\bar{{\beta}}_{2}\; , \qquad \delta {B}_2= 2\Re {\rm e}\,\beta_2 = \beta_2 +  \bar{{\beta}}_{2}\; . \qquad
\end{eqnarray}

The expression (\ref{bcA3=bSi}) is clearly gauge fixed and the residual gauge symmetry preserving this form of the complex 3--form potential
 is generated by the following transformations of complex linear and real linear superfields
\begin{eqnarray} \label{vSi=tL}
\delta {\Sigma}=\tilde{{\bb L}} + i{\bb L}\, , \qquad \delta \bar{{\Sigma}}= \tilde{{\bb L}}  -i{{\bb L}} \, , \qquad \delta L ={\bb L} \, . \qquad
\end{eqnarray}
The transformations 'parametrized' by the second real linear multiplet,  $\tilde{{\bb L}}$, leave invariant the real 3-form $C_3$ which enters the
supermembrane action. However, it is convenient to introduce the St\"uckelberg real linear multiplet superfield $\tilde{L}$ also for these  transformations,
\begin{eqnarray}\label{vL'=tL'}
 \delta \tilde{L} =\tilde{{\bb L}}{}\, , \qquad
\end{eqnarray}
so that $(\Sigma - \tilde{L}-iL)$ and its c.c. are gauge invariant.

The simplest action for the supergravity and double three form  matter supermultiplet(s) can be obtained by substituting
\eqref{cP=ImSi} for ${\cal P}$ and ($i/2$) ${\cal S}$ for  ${\cal Z}$ into \eqref{Ssg+1+=}.
However, with the above described St\"uckelberg realization of the two three form gauge symmetries we can write the action with a more general mass term, thus arriving at
\begin{eqnarray}\label{Ssg+2+=}
S_{sugra+matter}&=&-\frac 3 {4\kappa^2} \int d^8z \, E\, \Omega({\cal S},\bar{{\cal S}})
 -\frac 1 {2\kappa^2} \left( \int d^6\zeta_L \, {\cal E}\, {\cal W}({\cal S}) + c.c.\right) - \qquad \nonumber \\ && - {\frak {m}}^4 \int d^8z \, E\, \frac {\left({\Sigma}-\tilde{L}- iL \right) \left(\bar{\Sigma}-\tilde{L} + iL \right)}{({\cal S}\bar{{\cal S}})^{1/3}}
\; .  \qquad
\end{eqnarray}

Thus the coupling of open supermembrane to the simplest double three form matter and supergravity is described by the action \eqref{S=S+} with \eqref{Ssg+2+=},
and
  \begin{eqnarray}\label{Sp=2+1=Tc}
  S_{p=2}+S_{p=1}= \frac {T_2} 2 \int d^3 \xi \sqrt{|h|}\,|{\cal{S}}| +T_2\int\limits_{W^3} ({\cal A}_3+ \bar{\cal A}_3-H_3) + \qquad \nonumber \\
  +\frac {T_2} 4 \int\limits_{W^2=\partial W^3}d^2\sigma\sqrt{-\gamma} \, |{\Sigma}- \bar{\Sigma}- 2i{L}| \; . \qquad
\end{eqnarray}

\section{Open supermembrane, nonlinearly self-interacting double 3--form matter multiplets and supergravity }

In this section we would like to consider a coupling of open supermembrane and superstring at the boundary of supermembrane to the dynamical system of $n$ self-interacting double three form multiplets and supergravity \cite{Farakos:2017jme} which is of the type appearing in string compactifications. The interaction of closed supermembrane with such a system was studied in \cite{Bandos:2018gjp}.

Let us consider a  set of $(n+1)$  special chiral superfields ${\cal S}^I$, $I=0,1,...,n$ which are defined by a nonlinear interacting generalization of the above discussed \eqref{S=Si}. We describe them below in Eq. (\ref{SI=SiJ}) and now just state that each of them carry  two three form fields among their components.

Following \cite{Farakos:2017jme} and \cite{Bandos:2018gjp}, let us consider ${\cal S}^I$ as coordinates of a special K\"ahler manifold with holomorphic prepotential ${\cal G}({\cal S})$ homogeneous of order two,
\begin{eqnarray}\label{cG=hom2}
{\cal G}(w{\cal S})=w^2 {\cal G}({\cal S})
\, . \qquad
\end{eqnarray}
Then
\begin{eqnarray}\label{cGI:=}
{\cal G}_I({\cal S}) = \partial_I {\cal G}({\cal S})= {\cal G}_{IJ}({\cal S}) {\cal S}^J \qquad
\end{eqnarray} and \begin{eqnarray}\label{cGIJ:=}
  {\cal G}_{IJ}({\cal S}) := \partial_I\partial_J {\cal G}({\cal S})
\,  \qquad
\end{eqnarray}
are homogeneous of degrees one and zero, respectively.
We define our special chiral  superfields by \cite{Farakos:2017jme,Bandos:2018gjp}
\begin{eqnarray}\label{SI=SiJ}
{{\cal S}}^I &=& \frac 1 4 \left(\bar{{\cal D}}_{\dot\alpha}\bar{{\cal D}}{}^{\dot\alpha}- R\right) {\cal M}^{IJ}  (\Sigma_J -  \bar{{\Sigma}}_J) \; \qquad \nonumber \\
&=& \frac i 2 \left(\bar{{\cal D}}_{\dot\alpha}\bar{{\cal D}}{}^{\dot\alpha}- R\right) {\cal M}^{IJ}  \Im {\rm m}\Sigma_J \; , \qquad
\end{eqnarray}
where $\Sigma_J=  (\bar{{\Sigma}}_J)^*$ are complex linear superfields,
\begin{eqnarray}\label{DDSIi=bRSiI}
({\cal D}^\alpha {\cal D}_\alpha -\bar{R})\Sigma_J=0\; , \qquad \left(\bar{{\cal D}}_{\dot\alpha}\bar{{\cal D}}{}^{\dot\alpha}- R\right)  \bar{{\Sigma}}_J=0 \; , \qquad
\end{eqnarray}
and the real symmetric matrix ${\cal M}^{IJ}$ is the inverse of the imaginary part of \eqref{cGIJ:=},
\begin{eqnarray}\label{cMIJ:=}
{\cal M}^{IJ} {\cal M}_{JK}=\delta^I{}_K\; , \qquad {\cal M}_{IJ}:= \Im{\rm m}\, {\cal G}_{IJ}\;  .
\end{eqnarray}

Generically, the relation \eqref{SI=SiJ} is nonlinear, while in a particular case of  $ {\cal G}_{IJ}({\cal S})=i\delta_{IJ}$,
$ {\cal M}^{IJ}= \delta_{IJ}$ and it reduces to the set of $n+1$ independent relations (\ref{S=Si}).
Notice that the homogeneity of the holomorphic prepotential implies that
\begin{eqnarray}\label{GI=SiJ}
{{\cal G}}_I = {{\cal G}}_{IJ}({{\cal S}}) {{\cal S}}^J=
 \frac 1 4 \left(\bar{{\cal D}}_{\dot\alpha}\bar{{\cal D}}{}^{\dot\alpha}- R\right) {{\cal G}}_{IK} {\cal M}^{KJ}  (\Sigma_J -  \bar{{\Sigma}}_J) \qquad \nonumber \\  = \frac i 2 \left(\bar{{\cal D}}_{\dot\alpha}\bar{{\cal D}}{}^{\dot\alpha}- R\right) \Im {\rm m}\, ( {{\cal G}}_{IK} {\cal M}^{KJ}\Sigma_J )\; ,  \qquad
\end{eqnarray}
so that the composite chiral superfield  ${{\cal G}}_I({{\cal S}})$ in \eqref{cGI:=} is also special, of the type define in Eq. (\ref{S=Si}).

This observation makes manifest that the composite chiral superfield
 \begin{eqnarray}\label{cS=qISI-}
 {\cal S} = q_I {{\cal S}}^I-  p^I {{\cal G}}_{I}({{\cal S}}) \;  \qquad
\end{eqnarray}
with constants $q_I$ and $p^I$ is also special. Namely, it can be defined by equation (\ref{T2=bDbDP}) with
${\cal Z}= \frac i 2 {\cal S}$ and real prepotential given by a linear combination
\begin{eqnarray}\label{WZprepot1} \begin{aligned}
     {{\cal P}} & = q_I {{\cal P}}^I- p^I \tilde{{\cal P}}_I
\end{aligned}
\end{eqnarray}
of the composite  real superfields
 \begin{eqnarray}\label{realprepot1}
{{\cal P}}^I\equiv \Im {\rm m} ({{\cal M}}^{IJ}\Sigma_J )\; ,\qquad \tilde{{\cal P}}_I\equiv \Im {\rm m}({\cal G}_{IJ}{{\cal M}}^{JK}\Sigma_K)
\;  \qquad
\end{eqnarray}
which serve as prepotentials  for ${{\cal S}}^I$ and ${\cal G}_I({{\cal S}})$
(see (\ref{SI=SiJ}) and  (\ref{GI=SiJ})),
\begin{eqnarray}\label{defSG}
{{\cal S}}^I=\frac i 2(\bar{{\cal D}}^2-R){{\cal P}}^I\,,\qquad {\cal G}_I({{\cal S}})=\frac i 2(\bar{{\cal D}}^2-R)\tilde{{\cal P}}_I \, .
\end{eqnarray}

The composite superfield (\ref{cS=qISI-}) has been used in \cite{Bandos:2018gjp} to couple the closed supermembrane to double three form matter and supergravity.
Here we will be using it in the open supermembrane action (\ref{Sp=2:=}) setting
$-2i{\cal Z}\mapsto   {\cal S} = q_I {{\cal S}}^I-  p^I {{\cal G}}_{I}({{\cal S}})$ \eqref{cS=qISI-} and ${{\cal P}}\mapsto  q_I {{\cal P}}^I- p^I \tilde{{\cal P}}_I$ \eqref{WZprepot1}.
\if{}
$C_3\mapsto q_I {\cal C}_3^I - p^I  \tilde{{\cal C}}_{3I} $ where
$ {\cal C}_3^I$ and  $ \tilde{{\cal C}}_{3I} $ are calculated like in Eq. (\ref{C3=cP}) using the composite prepotentials
${{\cal P}}^I$ and $\tilde{{\cal P}}_I$ of Eq. (\ref{realprepot1}), respectively.
\fi

The supermembrane action in the background of nonlinearly self-interacting three form matter and supergravity reads
  \begin{eqnarray}\label{Sp=2=}
  S_{p=2}(q_I,p^I) = {1\over 2}\int d^3 \xi \sqrt{|\hat{g}|}\,| q_I {{\cal S}}^I-  p^I {{\cal G}}_{I}({{\cal S}})| +q_I \int_{W^3} {\cal C}_3^I - p^I  \int_{W^3}\tilde{{\cal C}}_{3I} \;  \qquad
\end{eqnarray}
where  $ {\cal C}_3^I$ and  $ \tilde{{\cal C}}_{3I} $ have the form of  (\ref{C3=cP}) with
${{\cal P}}^I$ and $\tilde{{\cal P}}_I$ from Eqs. (\ref{realprepot1}), ${\cal S}^I$ is given by (\ref{SI=SiJ}) and
${{\cal G}}_{I}({{\cal S}})$ is defined in (\ref{cGI:=}) and  has the form of  (\ref{GI=SiJ}) due to (\ref{cG=hom2}).

It is not difficult to check that the special chiral superfield (\ref{SI=SiJ}) does not change under the transformations
\begin{eqnarray} \label{vSiI=tLI}
\delta {\Sigma}_I=\tilde{{\bb L}}_I + {{\cal G}}_{IJ} {\bb L}^J\, , \qquad \delta \bar{{\Sigma}}_I=\tilde{{\bb L}}_I + \bar{{\cal G}}_{IJ} {\bb L}^J\, , \qquad
\end{eqnarray}
with real  linear superfields ${\bb L}^I$ and $\tilde{{\bb L}}_I$,
\begin{eqnarray}\label{DDbbLI=0=}
\left(\bar{{\cal D}}_{\dot\alpha}\bar{{\cal D}}{}^{\dot\alpha}- R\right){\bb L}^I=0=
\left({{\cal D}}^{\alpha}{{\cal D}}{}_{\alpha}- \bar{R}\right){\bb L}^I   \; , \qquad \left(\bar{{\cal D}}_{\dot\alpha}\bar{{\cal D}}{}^{\dot\alpha}- R\right)\tilde{{\bb L}}_I  =0=
\left({{\cal D}}^{\alpha}{{\cal D}}{}_{\alpha}- \bar{R}\right)\tilde{{\bb L}}_I   \; .
\end{eqnarray}
As in a simpler cases discussed in previous section, (\ref{vSiI=tLI}) generate a particular case of the gauge transformations of super-3-forms,
\begin{eqnarray}
 \label{vcA3=db2}
\delta {{\cal C}}^I_{3}=d\beta^I_2 \; , \qquad
\delta  \tilde{{\cal C}}_{3I} =d\tilde{{\beta}}_{2I}\;  \qquad
\end{eqnarray}
with $d\beta^I_2$ and $d\tilde{{\beta}}_{2I}$ expressed in terms of ${\bb L}^I $ and $\tilde{{\bb L}}_I$ as in (\ref{dal2=}).

Thus for  closed supermembrane the above action (\ref{Sp=2=}) is invariant under (\ref{vSiI=tLI}). To reach the same for an open supermembrane we should add to (\ref{Sp=2=}) with $\partial W^3\not = \emptyset $ the action of superstring the worldsheet of which is the boundary of the worldvolume, $W^2=\partial W^3$,
 \begin{eqnarray}\label{Sp=1=pq}
  S_{p=1}(q_I,p^I) &=& {1\over 2} \int\limits_{W^2}d^2\sigma\sqrt{- \gamma} \, |q_I({\cal P}^I-{L}^I) - p^I(\tilde{{\cal P}}_I-\tilde{L}_I)| - q_I\int\limits_{W^2} {B}^I_2 +  p^I\int\limits_{W^2} \tilde{B}_{2I}\; .   \qquad
\end{eqnarray}
In this boundary term of the supermembrane action  ${\cal P}^I$ and $\tilde{{\cal P}}_I$ are 'electric' and 'magnetic' parts of the composite real prepotentials \eqref{WZprepot1} defined by Eqs.  \eqref{realprepot1}, and $H^I_3=dB^I_2$ and $\tilde{H}_{3I}=d\tilde{B}_{2I}$ are expressed in terms of  the real linear superfields $L^I$ and $\tilde{L}_{I}$ as in (\ref{H3=dB2=}).

The sum of the open supermembrane and superstring actions \eqref{Sp=2=} and \eqref{Sp=1=pq} is invariant under the $\kappa$-symmetry
\eqref{ikappaEA=} provided  the parameters obey
\begin{eqnarray}\label{kappa=qpGbk}
 \kappa_\alpha = - i \frac { q_I {{\cal S}}^I-  p^I {{\cal G}}_{I}({{\cal S}})}  {| q_I {{\cal S}}^I-  p^I {{\cal G}}_{I}({{\cal S}})|} \; \Gamma_{\alpha\dot\alpha}\bar{\kappa}{}^{\dot\alpha} \; , \qquad
 \bar{\kappa}{}_{\dot\alpha}= - i \frac { q_I \bar{{\cal S}}^I-  p^I \bar{{\cal G}}_{I}(\bar{{\cal S}})}  {| q_I {{\cal S}}^I-  p^I {{\cal G}}_{I}({{\cal S}})|} \;  \kappa^\alpha \Gamma_{\alpha\dot\alpha} \; , \qquad
 \end{eqnarray}
and
\begin{eqnarray}\label{kappa-str-qp}
\kappa_\alpha = \frac {q_I({\cal P}^I-{L}^I) - p^I(\tilde{{\cal P}}_I-\tilde{L}_I)}{|q_I({\cal P}^I-{L}^I) - p^I(\tilde{{\cal P}}_I-\tilde{L}_I)|} \,  P_\alpha{}^\beta \kappa_\beta\, , \qquad \nonumber \\
\bar\kappa_{\dot\alpha} = \, \frac {q_I({\cal P}^I-{L}^I) - p^I(\tilde{{\cal P}}_I-\tilde{L}_I)}{|q_I({\cal P}^I-{L}^I) - p^I(\tilde{{\cal P}}_I-\tilde{L}_I)|} \,  \bar P_{\dot\alpha}{}^{\dot\beta}\bar \kappa_{\dot\beta}\, , \qquad
\end{eqnarray}
with the projectors defined in \eqref{G=} and \eqref{proj-str}.

The sum of the open supermembrane and superstring actions is also invariant under the gauge symmetry (\ref{vSiI=tLI}) supplemented by
the St\"uckelberg--type transformations of the real linear superfields:
\begin{eqnarray} \label{vSiI=bbLI}
\delta {\Sigma}_I=\tilde{{\bb L}}_I + {{\cal G}}_{IJ} {\bb L}^J\, , \qquad \delta L^I=  {\bb L}^I\, , \qquad \delta \tilde{L}_I =\tilde{{\bb L}}_I \, . \qquad
\end{eqnarray}
These leave invariant the chiral superfields (\ref{SI=SiJ}) as well as the combinations
\begin{eqnarray} \label{SiI-=inv}
 {\Sigma}_I- \tilde{{L}}_I - {{\cal G}}_{IJ} {L}^J\, . \qquad
\end{eqnarray}

A quite general action for nonlinearly self interacting system of the 3-form multiplets and supergravity, of a kind which appear in string compactifications, with a spontaneously broken (realized by St\"uckelberg mechanism) 3-form gauge symmetry reads
\begin{eqnarray}\label{Ssg+2n+=}
S_{sugra+matter}&=&-\frac 3 {4\kappa^2} \int d^8z \, E\, \Omega({\cal S}^I,\bar{{\cal S}}^I)
 -\frac 1 {2\kappa^2} \left( \int d^6\zeta_L \, {\cal E}\, {\cal W}({\cal S}^I) + c.c.\right) - \qquad \nonumber \\ && - c_2 \int d^8z \, E\, \frac {{\cal M}^{IJ}\left(
 {\Sigma}_I- \tilde{{L}}_I - \bar{{\cal G}}_{IK} {L}^K\right) \left(\bar{\Sigma}_J-\tilde{L}_J - {{\cal G}}_{IL} {L}^L\right)}{({\cal S}^{P}{\cal M}_{PQ}\bar{{\cal S}}{}^{Q})^{\frac 1 3}}
\; .  \qquad
\end{eqnarray}
This can be included as matter plus supergravity part in the interacting action \eqref{S=S+} together with the open supermembrane and superstring at the boundary of supermembrane actions \eqref{Sp=2=} and \eqref{Sp=1=pq}.
Such an action will be invariant under the $Sp(2n+2|{\bb Z})$ symmetry, characteristic for string compactifications, provided
$\Omega({\cal S}^I,\bar{{\cal S}}^I)$ and ${\cal W}({\cal S}^I) $ are invariant and the supermembrane charges $(p^I,q_I)$ in  \eqref{Sp=2=} and \eqref{Sp=1=pq} are transformed as symplectic vector.
Actually the quantization of these charges  breaks the possible  $Sp(2n+2|{\bb R})$ symmetry of the supergravity plus  matter (super)field system to its discrete subgroup $Sp(2n+2|{\bb Z})$ (see \cite{Bandos:2018gjp} for a discussion of quantization of $(p^I,q_I)$ without a reference to the higher dimensional origin of the D=4 domain wall system).

\section{Conclusion}

In this paper we present the actions describing the interacting dynamical system of open supermembrane, quite generic 3-form matter and supergravity. The similar interacting system containing closed supermembrane was studied in \cite{Bandos:2018gjp}. The action of open supermembrane in Veneziano-Yankelovich effective theory of ${\cal N}=1$ SYM  was discussed in \cite{Bandos:2019qok}.

We have begun by writing the action for the most general interacting system of open supermembrane, single three form matter and supergravity, which serves  as a master case for the derivation of the actions for more complicated systems. A particular case of this action describes the  interacting system of open supermembrane and single three form supergravity.

Then we describe the interacting actions for the open supermembrane interacting with double 3-form matter multiplet and supergravity. Finally, we present the action for a quite generic system of open supermembrane, a number of nonlinearly interacting double three form multiplets and supergravity which has a  special K\"ahler structure and possesses an invariance under symplectic transformations; the dynamical field theoretical systems of such a type appear in string theory compactifications.

To preserve the $\kappa$--symmetry, the  open supermembrane action should include a boundary term which have a natural interpretation of the action for superstring at the end of supermembrane. In contrast with the $\kappa$-symmetry, the three form gauge symmetry is broken when the  supermembrane is open. This breaking can be interpreted as spontaneous and the 3-form gauge symmetry can be formally maintained with the use of St\"uckelberg mechanism. To this end we have to introduce the pure gauge real linear superfield(s) $L$ ($L^\Lambda=(L^I, \tilde{L}_I)$) and the Wess--Zumino term of the action of superstring at the end of supermembrane is constructed with the use of this (these) supermultiplet(s). The gauge fixed version of this superstring action contains the Nambu-Goto-type term only and the tension of superstring at the end of supermembrane is expressed in terms of real  prepotential superfield(s) ${\cal P}$ (${\cal P}^\Lambda$), either fundamental or composite, of the special chiral superfield(s) describing 3-form matter supermultiplet(s) and/or conformal compensator of the 3-form supergravity interacting with the open supermembrane.

As 3-form gauge symmetry is broken due to the presence of open supermembrane,  or realized by St\"uckelberg mechanism,  we find natural to consider also the  terms breaking the 3-form gauge symmetry in the supergravity plus matter part of the action: the generalized mass terms.


The inclusion of the actions of open supermembranes  with closed strings at the boundary of supermembranes
into effective field theories (EFT) of string theory flux compactifications is inline with the completeness conjecture \cite{Polchinski:2003bq}\footnote{Actually in \cite{Polchinski:2003bq} Polchinski proposed two completeness principles, the second of which, most relevant for our discussion, states that ''in any fully unified theory, for every gauge field there will exist electric and magnetic sources"  obeying the Dirac quantization conditions $eg=2\pi n$  "with the minimum relative Dirac quantum
$n = 1$
(more precisely, the lattice of electric and magnetic charges
is maximal)" \cite{Polchinski:2003bq}. The first of the completeness principles relate any charge quantization to the existence of magnetic monopoles.}. The EFT action of such a type can be obtained e.g. from type IIB compactifications on wrapped Calabi-Yau manifolds  of the dynamical systems including the  networks of D7-,  D5- and D3-branes  discussed in \cite{Evslin:2007ti} and \cite{BerasaluceGonzalez:2012zn}. In recent \cite{Lanza:2019xxg}, appearing on the net slightly after the first version of the present paper, the actions for network of open supermembranes and strings were obtained independently and completed by an interesting  actions for spacetime filling 3-branes which are given by a  Wess--Zumino-type  terms only and do not break any supersymmetry.


To resume, we believe that the actions presented in this paper as well as its generalizations discussed in \cite{Lanza:2019xxg}
will be useful to construct the effective actions for phenomenologically interesting models of string theory compactifications with open branes and branes at the boundary of open branes.
The natural next step in development of our formalism  is to obtain equations of motion for 3-form matter and supergravity from our action and to search for their solution describing open supermembrane systems and supermembrane junctions.

\subsection*{Acknowledgements}
The author is thankful to  Sergei Kuzenko and especially to Dima Sorokin for useful conversations and for collaboration at early stages of this work which was supported in part by the Spanish MINECO/FEDER (ERDF EU)  grant PGC2018-095205-B-I00, by the Basque Government Grant IT-979-16, and by the Basque Country University program UFI 11/55.

\appendix
\section{Torsion constraints of minimal supergravity }
\setcounter{equation}0
\renewcommand{\theequation}{A.\arabic{equation}}

Our notations are those of \cite{Bandos:2002bx,Bandos:2012da,Bandos:2012gz}; they are close but not identical to that of \cite{Wess:1992cp}. In particular we use the mostly minus metric conventions, $\eta_{ab}=diag (+,-,-,-)$ and the set of our superspace constraints contains
$R_{\alpha\dot\beta}{}^{cd}=0$ instead of $T_{ab}{}^c=0$ in  \cite{Wess:1992cp}.

The superspace constraints of minimal supergravity and their consequences can be collected in the following expressions for superspace torsion and curvature 2-forms
           \begin{eqnarray}\label{4WTa=} && T^a
= {\cal D}E^a=- 2i\sigma^a_{\alpha\dot{\alpha}} E^\alpha \wedge \bar{E}^{\dot{\alpha}} -{1\over 8} E^b \wedge E^c
\varepsilon^a{}_{bcd} G^d \; ,  \hspace{4.0cm} \\ \label{4WTal=} && T^{\alpha} = {\cal D}E^{\alpha}  = {i\over 8} E^c \wedge E^{\beta}
(\sigma_c\tilde{\sigma}_d)_{\beta} {}^{\alpha} G^d   -{i\over 8} E^c
\wedge \bar{E}^{\dot{\beta}} \epsilon^{\alpha\beta}\sigma_{c\beta\dot{\beta}}R +
 {1\over 2} E^c \wedge E^b \; T_{bc}{}^{\alpha} \; , \\
\label{4WTdA=} && T^{\dot{\alpha}}  = {\cal D} E^{\dot{\alpha}}= {i\over 8} E^c \wedge E^{\beta} \epsilon^{\dot{\alpha}\dot{\beta}}
\sigma_{c\beta\dot{\beta}} \bar{R}  -{i\over 8} E^c \wedge
\bar{E}^{\dot{\beta}} (\tilde{\sigma}_d\sigma_c)^{\dot{\alpha}}{}_{\dot{\beta}} \, G^d +
 {1\over 2} E^c \wedge E^b \; T_{bc}{}^{\dot{\alpha}}\; ,
\end{eqnarray}
\begin{eqnarray}\label{4WR=def}
R^{ab} = {1\over 2} R^{\alpha\beta}
(\sigma^a\tilde{\sigma}^b)_{\alpha\beta}
- {1\over 2} R^{\dot{\alpha}\dot{\beta}}
(\tilde{\sigma}^a\sigma^b)_{\dot{\alpha}\dot{\beta}} \; ,
\end{eqnarray}
\begin{eqnarray}\label{4WR=}  R^{\alpha\beta} &\equiv &   {1 \over 4} R^{ab} \sigma_{ab}{}^{\alpha\beta}=
-{1\over 2} E^\alpha \wedge E^\beta \bar{R} -{i\over 8} E^c \wedge E^{(\alpha}\,
\tilde{\sigma}_c{}^{\dot{\gamma}\beta)} \bar{{\cal D}}_{\dot{\gamma}}\bar{R}+ \qquad \\ && + {i\over 8}
E^c \wedge E^{\gamma} (\sigma_c\tilde{\sigma}_d)_{\gamma}{}^{(\beta} {\cal D}^{\alpha)} G^d -  \nonumber
{i\over 8}  E^c \wedge \bar{E}^{\dot{\beta}} \sigma_{c\gamma\dot{\beta}} W^{\alpha\beta\gamma} +
{1\over 2} E^d \wedge E^c R_{cd}{}^{\alpha\beta} \; ,
\end{eqnarray}
$R^{\dot{\alpha}\dot{\beta}}=
(R^{{\alpha}{\beta}})^*$,
and in the following  equations for main  superfields
\begin{eqnarray} \label{chR} && {\cal D}_\alpha \bar{R}=0\;
, \qquad \bar{{\cal D}}_{\dot{\alpha}} {R}=0\; ,
 \\
\label{DG=DR} && \bar{{\cal
D}}^{\dot{\alpha}}G_{{\alpha}\dot{\alpha}}= -{\cal D}_{\alpha} R \; , \qquad {{\cal
D}}^{{\alpha}}G_{{\alpha}\dot{\alpha}}= -\bar{{\cal D}}_{\dot{\alpha}} \bar{R} \; , \qquad \\
\label{chW} && \bar{{\cal D}}_{\dot{\alpha}} W^{\alpha\beta\gamma}= 0\; , \qquad {{\cal D}}_{{\alpha}}
\bar{W}^{\dot{\alpha}\dot{\beta}\dot{\gamma}}= 0\;, \qquad
\\ \label{DW=DG}
&& {{\cal D}}_{{\gamma}}W^{{\alpha}{\beta}{\gamma}}= \bar{{\cal D}}_{\dot{\gamma}} {{\cal
D}}^{({\alpha}}G^{{\beta})\dot{\gamma}} \; , \qquad \bar{{\cal D}}_{\dot{\gamma}}
\bar{W}^{\dot{\alpha}\dot{\beta}\dot{\gamma}} = {{\cal D}}_{{\gamma}} \bar{{\cal D}}^{(\dot{\alpha}|}
G^{{\gamma}|\dot{\beta})} \; . \qquad \end{eqnarray}

As a consequence, the superalgebra of superspace covariant derivatives ${\cal D}_A$  (\ref{calD=})
in the case of minimal supergravity contains the following anticommutators
 of  fermionic covariant derivatives
    \begin{eqnarray}\label{cDcbD=Da}
   \{ {\cal D}_\alpha,  \bar{{\cal D}}_{\dot{\alpha}}\} &=& 2i \sigma^a_{\alpha\dot{\alpha}} {\cal D}_a \; ,
  \\
\label{DD=} &&
\{ {\cal D}_\alpha , {\cal D}_\beta \} \; V_\gamma = - \bar{R}
\epsilon_{\gamma (\alpha } V_{\beta)}\; ,  \\ \label{bDbD=}  &&  \{ \bar{\cal D}_{\dot\alpha} , \bar{\cal D}_{\dot \beta} \} \; V_{\dot \gamma}= \; {R}
\epsilon_{\dot{\gamma} (\dot{\alpha} } V_{\dot{\beta} )}\; . \end{eqnarray}

\section{Super-Weyl symmetry of the minimal  supergravity constraints  }
\setcounter{equation}0
\renewcommand{\theequation}{B.\arabic{equation}}

The super-Weyl transformations with covariantly chiral superfield parameter $\Upsilon$,
\begin{eqnarray}
\label{bDUp=0} && \bar{\cal D}_{\dot{\alpha}}\Upsilon=0\; , \qquad
\end{eqnarray}
 which leave invariant minimal (and 3-form) supergravity constraints, as well as, for instance,  the action \eqref{S=Ssg+Sp2} with \eqref{Ssg=} and \eqref{Sp=2:=},
are defined by \cite{Howe:1978km}
\begin{eqnarray}
\label{supW=m4Db}
E^a &\mapsto & \tilde{E}{}^a= e^{\Upsilon+\bar{\Upsilon}} E^a \; , \qquad  \\
\label{supW=m4Df}
E^{\alpha} &\mapsto & \tilde{E}{}^{\alpha}= e^{2\bar{\Upsilon}-{\Upsilon}} \left(E^{\alpha} - \frac{i}{2} E^a \bar{\cal D}_{\dot{\alpha}} \bar{\Upsilon} \tilde{\sigma}{}_a^{\dot{\alpha}{\alpha}} \right)\; , \qquad \\
\label{supW=m4Dbf}
\bar{E}{}^{\dot\alpha} &\mapsto & \tilde{\bar{E}}{}^{\dot\alpha}= e^{2{\Upsilon}-\bar{\Upsilon}} \left(\bar{E}{}^{\dot\alpha} - \frac{i}{2} E^a \tilde{\sigma}{}_a^{\dot{\alpha}{\alpha}} {\cal D}_\alpha {\Upsilon} \right)\; , \qquad \\
\label{supW=cZ} {\cal Z}  &\mapsto &  e^{-6{\Upsilon}} {\cal Z} \; . \qquad
\end{eqnarray}
It is useful to notice that under these transformations
\begin{eqnarray}
\label{supW=bDbD-R}
(\bar{{\cal D}}\bar{{\cal D}} -R)... &\mapsto & e^{-4{\Upsilon}}(\bar{{\cal D}}\bar{{\cal D}} -R)e^{2\bar{\Upsilon}}... \; , \qquad \\
\label{supW=cP}  {\cal P}  &\mapsto &  e^{-2\Upsilon-2\bar{\Upsilon}} {\cal P} \; , \qquad  \\
\label{supW=cE} {E} &\mapsto & e^{2\Upsilon+2\bar{\Upsilon}}{E}\; , \qquad  \\
\label{supW=cE} {\cal E} &\mapsto & e^{6{\Upsilon}}{\cal E}\; , \qquad \\
\label{supW=cW} {\cal W} &\mapsto & e^{-6{\Upsilon}}{\cal W}\; . \qquad
\end{eqnarray}

\section{Useful equations on supermembrane worldvolume}
\setcounter{equation}0
\renewcommand{\theequation}{C.\arabic{equation}}

 The orientation of the volume element of $W^3$ is defined by
  \begin{eqnarray}\label{d3xi=}
d\xi^m\wedge d\xi^n\wedge d\xi^k= d^3\xi \epsilon^{mnk}\; , \qquad  \epsilon^{012}=1\; , \qquad
 \end{eqnarray}
 The invariant measure with induced metric on $W^3$ can be written in the following equivalent form
  \begin{eqnarray}\label{d3xisqh=}
   d^3\xi\sqrt{h}= - \frac 1 3  *E_a\wedge E^a\; , \qquad  d^3\xi\delta \sqrt{h}= -  * E_a\wedge \delta E^a\; \qquad
 \end{eqnarray}
 with the Hodge star operation defined by
 \begin{eqnarray}\label{*Ea=}
    *E^a= \frac 1 2 d\xi^n\wedge d\xi^m  \sqrt{h}  \epsilon_{mnk}h^{kl}E_{l}^a\; . \qquad
 \end{eqnarray}
 The definition of the $\kappa$--symmetry projector in \eqref{G=} implies
  \begin{eqnarray}\label{d3xiG=}
  d^3\xi\sqrt{h}\Gamma_{\alpha\dot\alpha}=\frac i {3!} \sigma^a_{\alpha\dot\alpha} \epsilon_{abcd} E^b\wedge E^c\wedge E^d \; ,\qquad \\
  \label{EaEfstG=} *_3 E_a\wedge E^\beta (\sigma^a\tilde{\Gamma})_\beta{}^\alpha = - \frac 1 2  E^b\wedge E^c \wedge  E^\beta (\sigma_{bc})_\beta{}^\alpha
\; . \qquad
 \end{eqnarray}

\end{document}